**Graphical Abstract:**

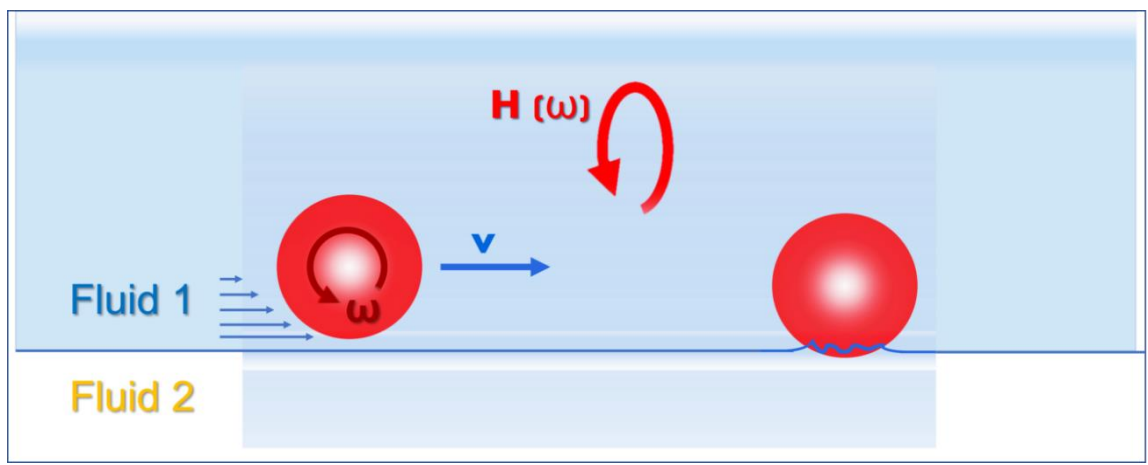

**Caption:** Microspheres forced to rotate in the vicinity of a fluid interface exhibit a roto-translational hydrodynamic mechanism that is strongly hampered by capillary torques as soon as the particles protrude from the interface. Under these conditions, the time evolution in the number of moving spheres provides a direct description of the adsorption kinetics, while microscopy monitoring of particle deceleration informs about the adsorption dynamics.

# Rotating Micro-Spheres for Adsorption Monitoring at a Fluid Interface

Martín-Roca J.[1,2], Jiménez M.[3], Ortega F. [3,4], Calero C.[5,6], Valeriani Ch. [1,2], Rubio R.G.[3,4] Martínez-Pedrero F.[3,*]

[1]Departamento de Estructura de la Materia, Física Térmica y Electrónica, Universidad Complutense de Madrid, 28040 Madrid, Spain

[2]GISC-Grupo Interdisciplinar de Sistemas Complejos, 28040 Madrid, Spain

[3]Departamento de Química Física, Universidad Complutense de Madrid, Avda. Complutense s/n, Madrid, 28040, Spain.

[4]Inst. Pluridisciplinar, Universidad Complutense de Madrid, Paseo Juan 23,1, E-28040 Madrid, Spain.

[5]Departament de Física de la Matèria Condensada, Universitat de Barcelona, 08028 Barcelona, Spain.

[6]Institut de Nanociència i Nanotecnologia, IN2UB, Universitat de Barcelona, 08028 Barcelona, Spain.

*fernandm@ucm.es

**Abstract**

**Hypothesis:** A broad range of phenomena, such as emulsification and emulsion stability, foam formation or liquid evaporation, are closely related to the dynamics of adsorbing colloidal particles. Elucidation of the mechanisms implied is key to a correct design of many different types of materials.

**Experiments:** Microspheres forced to rotate near a fluid interface exhibit a roto-translational hydrodynamic mechanism that is hindered by capillary torques as soon as the particles protrude the interface. Under these conditions, the time evolution in the ratio of moving spheres provides a direct description of the adsorption kinetics, while microscopy monitoring of particle acceleration\deceleration informs about the adsorption\desorption dynamics. In this work, the proposed strategy is applied at an air/water interface loaded with spherical magnetic particles negatively charged, forced to rotate by the action of a rotating magnetic field.

**Findings:** The proposed method enables the adsorption/desorption dynamics to be followed during the earliest phase of the process, when desorption of a small fraction of particles is detected, as well as to estimate approximated values of the adsorption/desorption constants. The results obtained show that the addition of a monovalent salt or a cationic (anionic) surfactant promotes (inhibits) both adsorption and formation of permanent bonds between particles.

## 1. Introduction

Colloidal micro-nanoparticles are used today in the stabilization of foams and Pickering emulsions, in a wide variety of industrial, environmental and technological processes, including oil recovery, drug delivery, catalysis, development of microfluidic systems for biotechnology applications, membrane technology, as well as in the suppression of the coffee ring effect [1-5]. In all these processes, the stability of the particle-laden fluid interfaces is determined by different factors, such as the contact angle, the adsorption kinetics, the particle dynamics at the interfaces, the competition for adsorption with other surface-active agents, and the shear and dilational rheological properties [2,6].

In diluted conditions, the adsorption process can be modeled by the standard Langmuir adsorption isotherm, which assumes that the adsorbent surface is only covered with a monolayer of adsorbate molecules, and that there is no interaction between them on the adsorbent surface. The model, too simplified for most real systems, predicts an asymptotic approach to full adsorption saturation on the adsorbent surface, and provides a useful conceptual frame for rationalizing the adsorption process [7]. The adsorption kinetics of colloids at a fluid interface has been traditionally characterized by means of indirect techniques such as—among others— neutron reflectometry and Wilhelmy or pendant drop tensiometry, widely used to investigate the adsorption kinetics of surfactants [8,9]. In most of the experimental conditions, the diffusive bulk transport toward the interface and the adsorption/desorption processes are coupled [7,10]. Diffusion is only negligible when the characteristic volume of the media containing the particles is smaller than $V^* = (D/(k'_{ads}\Xi^\infty))^3$, where $D$ is the diffusion coefficient, $k'_{ads}$ the adsorption constant, and $\Xi^\infty$ the maximal interfacial coverage. Since the kinetic rate coefficients are not measurable when the rate of kinetics is much faster than the rate of diffusion, many previous studies have used increasing bulk concentrations, microfluidic tensiometers, convection and interface curvature to change the dominant mechanism from diffusive to kinetic, and extend the range of measurable kinetic processes [7,11]. Another obstacle with this approach is that the surface coverage by the microparticles, $\Xi(t)$, is not obtained directly in these techniques, so $\Xi(t)$ needs to connect with the surface pressure through an appropriate model, which accounts for the specific nature of the system [8].

In this area of research, there remains the problem of discerning whether particles are actually adsorbed at the fluid interface or not [6]. Even when the binding of the particle to the interface corresponds to a local free energy minimum, repulsion between the particles and the fluid interface can cause adsorption to be kinetically arrested and require external activation, which is usually induced by sonication, mechanical shaking, or the addition of salt, proteins, or surfactants [8,9]. The surface electric charge of the particles, responsible in many cases for the stability of the colloidal suspension, promotes the electrostatic repulsion with both the charged interface and the image charge arising when the two fluids composing the interface have different dielectric constants. The repulsion between the particles and the interface can also be due to other electro-confining effects, in which the electrical double layer of the particles approaching the boundary is deformed [8,12]. For very short timescales, the adsorption usually is limited by diffusion, so the surface tension $\gamma$ initially decays with $t^{1/2}$, while at longer times the repulsion between the particles and the interface imposes an exponential relaxation of $\gamma$ [9,13,14]. These two kinetics are implicit in the model developed by Ward and Torday [10], and more recently by Feinerman et al. [15]. The presence of salts and oppositely charged surfactant molecules partially screens the electrostatic repulsion, allowing particles to protrude the interface, but also promotes particle aggregation. The latter can become a problem, as extensive flocculation leads to the formation of large flocs that turn out to be ineffective for emulsion stabilization [1]. In parallel, oppositely charged surfactant molecules hydrophobize the surface of the particles, increasing their tendency to adsorb on the fluid-fluid interface, while all surface-active agents compete with the particles for adsorption [2,16].

If the particles overcome the repulsions with the interface, then they are able to initiate the process that, in most cases, ends with adsorption of the particle at the equilibrium position, where the fluid interface and the tangent to the contact line form the equilibrium contact angle $\theta_{eq}$. In systems in which the particles are partially hydrophilic/hydrophobic, particle adsorption is thermodynamically favorable, since the presence of particles at the interface reduces the contact area between the two phases, and thus the surface energy. If the adsorbing particle is a smooth sphere, small enough so that its weight can be neglected, its equilibrium position with respect to the fluid interface can be determined by minimizing the sum of three interfacial energies, those corresponding to the water-air $\gamma_{aw}$, water-particle $\gamma_{pw}$ and air-particle interfaces $\gamma_{ap}$ [17]. For perfectly smooth spherical particles with size above the nanoscale, $\theta_{eq}$ is related to the wettability of the particle through the equilibrium of forces at the contact line, leading to Young's equation:

$$\cos\theta_{eq} = \frac{\gamma_{ap} - \gamma_{pw}}{\gamma_{aw}}, \tag{1}$$

where the line tension is neglected, as well as the influence of Derjaguin forces near the three-phase contact line [18]. Besides, the energy reduction due to particle adsorption is given by:

$$\Delta E = -\pi a^2 \gamma (1 - |\cos\theta_{eq}|)^2. \tag{2}$$

If the value of the interfacial tension is high, the particle size is in the micrometer range and $\theta_{eq}$ is in the range between 0.1 and 3.0 rads, then adsorption reduces the surface energy more than a hundred times the thermal energy. As a consequence, the interfacial adsorption of the particles is thought to be virtually irreversible -in contrast to the dynamic adsorption/desorption process depicted by surfactant molecules, that are easily expelled toward the bulk phase by thermal energy fluctuations- and fairly low particle concentrations are enough to stabilize foams and emulsions [1]. In addition, the out of plane motion of the adsorbed particles, both translation and rotation, become strongly hindered [19].

Despite significant advances in the field, understanding the fundamental aspects of particle adsorption/desorption at fluid interfaces, such as the dynamics of binding, is elusive using large-volume methods. In recent years, the community has realized that this task requires of experiments that monitor and control the motion of the particles during the process [2,20-25]. For example, not many years ago, most theoretical and experimental studies implicitly assumed that particles almost immediately reach their equilibrium position within the interface. This assumption was based on the large reduction in surface energy linked to the adsorption process. However, using holographic and AFM techniques, it has been shown that during adsorption the particles do not reach the equilibrium contact angle as quickly. In fact, the whole process may take months, during which the particles cause a sudden rupture of the interface, followed by a slow relaxation, logarithmic in time [26-28]. The slow wetting dynamics, driven by a capillary force, agrees with an Arrhenius-type dynamics describing the thermally activated contact line "jump" on the nanometric heterogeneities present on the particle surface. Also recently, Boniello et al. [29] demonstrated, by using interferometric techniques, that the diffusive motion of an adsorbed colloidal particle at an air-water interface is slower than expected if only the viscosity of the participating fluids is considered. According to these authors, the observed behavior is due to the fact that the surface of the colloidal particles is heterogeneous at the nanometer scale, which induces the deformation of the anchored contact line. The fluctuation of the deformed three-phase line between surface irregularities is, according to the fluctuation-dissipation theorem, strictly linked to the occurrence of an effective interfacial viscosity, responsible for the slowing down of the translational and rotational diffusive processes.

To further investigate the dynamics and the adsorption/desorption mechanism of the colloids, as well as on the influence that the presence of different additives (monovalent electrolytes and anionic or cationic surfactants) has on the process, we introduce a direct, simple, and powerful method. In this method, we use magnetic microparticles as excitable probes for in situ monitoring of the adsorption process and perform a systematic study, using video-microscopy techniques, where we analyze in real time the individual and collective dynamics of micrometric particles adsorbing on fluid interfaces. In the experiments, particle adsorption is detected when the coupled rotational-translational motion exhibited by the excited magnetic particles is slowed down by the action of the capillary torque, which comes into play as soon as the particles pierce the fluid interface. In parallel, we have applied an alternative strategy used to simultaneously characterize the influence that different additives can have on the colloidal stability of the system. Electrophoretic mobility experiments were used to characterize the surface electrical properties of the particles.

2. **Methods and Materials**

**Superparamagnetic particles.** In this work we use a commercial aqueous suspension of superparamagnetic particles, Dynabeads® M-270 (Invitrogen). The particles are spherical, with a radius $a = 1.4$ μm, and monodisperse, with a variation coefficient CV <

3%. The beads are composed of a polystyrene matrix doped with nanometric superparamagnetic grains of iron oxides ($Fe_2O_3$ and $Fe_3O_4$), have a density $\rho_{particle}$ = 1.8 g/cm³, and are stabilized with carboxylic groups at the surface. The equipment used to characterize the mobility of the particles is the Zetasizer Nano model ZS from Malvern Instruments (USA), with a 632.8 nm He-Ne laser. All the mobility measurements were performed after thermostatization of the DTS1070 cell at a temperature of 25°C. In the assessment of the equilibrium contact angle, we have also used a commercial aqueous suspension of smaller superparamagnetic microspheres, Dynabeads® MyOne, supplied by Invitrogen (USA), 1.0 µm in diameter, monodisperse (CV < 5%), with identical composition and stabilized with carboxylic groups. Before preparing them at the required experimental conditions, the commercial colloidal suspensions are washed twice in pure water.

**Fluid air/water interface.** In our experiment, 200 µl of the aqueous colloidal dispersion, 0.02 %w, is placed on a glass slide with a 1.5 mm radius hole, where the drop is held by capillarity. After a few minutes, all the particles placed above the orifice are deposited by gravity on the slightly curved water-air interface, where they remain for the duration of the experiment (sketch in Figure 1). In pure water, the Bond number of the particles, $B_{particle} = \frac{(\rho_{particle} - \rho_{water})ga^2}{\gamma_{water}} = 10^{-6}$, is close to zero. Here, $g$ is the gravitational acceleration and $\gamma_{water}$ the water surface tension. The low particle concentration minimizes the interaction between the particles.

The excess droplet at the top serves as a water reservoir to prevent the interface from drying out during the process. A CCD camera (Edmund EO1312M, USA) is connected to an optical microscope (Nikon Eclipse 80-I with a working distance of 50×objective, 0.25 NA, Japon) that allows for imaging the fluid interface in the hole at single-particle resolution, at 64 frames s$^{-1}$. The interface curvature is small enough so that all settled particles are detected in the focus plane of the microscope. The chosen frame rate value allows to record videos during the long time necessary to detect the adsorption processes, while permitting to compute the particle velocity during the whole process. The FindMaxima routine from the public domain ImageJ is used to determine the position of the center of individual colloidal spheres in a sequence of frames. Each studied trajectory is composed of $N$ frames, always between 1000 and 3000. To calculate the velocity of the particles at point $M + 1$, the particle coordinates are fitted by a linear function in the frame range between 1 to $M$. Then, the value of the velocity at $M + 1 + dM$ is calculated in the same way from the coordinates in the range between $1 + dM$ and $M + dM$, and so on for the whole trajectory. If $dM < M$ then there is overlap between the different sets of points. On the other hand, the $M$ initial and final values of the velocity are not calculated. As $M = 10$ and $dM = 5$ are much smaller than $N$, the method applied does not significantly reduce the information collected.

**Rotating and constant magnetic field.** The sample is placed in the center of a set of coils made from insulated copper wire, used to generate the magnetic fields: a pair of coils oriented along the *x*-axis, facing each other at a distance equal to their diameter, and a third coil oriented along the *z*-axis, all placed in the proximities of the microscope stage. The signal feeding the coils is created with a signal generator (National Instruments 9269, USA) connected to three current amplifiers (KEPCO, bipolar power supply/operational amplifier, USA). The field applied on the sample is sufficiently homogeneous so as not to observe magnetic forces proportional to the gradient of the external field, which would drag particles through the medium. By modulating the current amplitude and frequency, different magnetic fields can be applied. Specifically, in this work, an elliptically polarized field has been applied in the *xz* plane, which is obtained as the sum of a sinusoidal signal in the *z*-axis $H_z = H_{z0} sin(2\pi f t)$ and a cosine one in the *x*-axis $H_x = H_{x0} cos(2\pi f t)$, both at the same frequency *f*. Here, $H_{z0}$ and $H_{x0}$ are the amplitudes of the applied field strengths along the corresponding axes. The elliptic field strength is given by $H_0 = \sqrt{\frac{H_{z0}^2 + H_{x0}^2}{2}}$, while the ellipticity of the field is characterized by the ellipticity parameter, defined by $\beta = \frac{H_{z0}^2 - H_{x0}^2}{H_{z0}^2 + H_{x0}^2}$. In experiments where the rotating field is applied uninterruptedly, the direction of rotation is changed periodically. The change of the direction of rotation is imposed to facilitate the tracking of the particle trajectory for as long as possible, without the particles leaving the field of view of the microscope.

**Contact Angle.** The contact angle of the particles Dynabeads® M-270 and Dynabeads® MyOne Carboxylic Acid with the water/air interface is small due to a high negative net charge at neutral pH, so after adsorbing at the fluid interface the particles remain almost immersed in the aqueous phase [30]. By applying the gel trapping technique (GTT) [24], we found that the equilibrium contact angle of the particles $\theta_{eq}$ is always smaller than 30º [19]. Since this technique requires substantial changes in the temperature of the studied system, as well as the use of gelling polymers, we decided to confirm the small value of the equilibrium contact angle via an alternative strategy. In this strategy, we studied the response of an adsorbed hetero-dimer -formed by the two different sized magnetic particles, Dynabeads® M-270 and Dynabeads® MyOne Carboxylic Acid- to the reorientation of a constant field, gradually tilted out of the confining boundary. It seems reasonable to assume that the contact angle of the two particles is identical, since the particles are formed from the same material and stabilized with carboxylic groups. In these conditions, the contact angle was obtained from the critical tilt angle of the field at which the confined particles pass from repulsive to attractive configurations, where the particles composing the dimer slowly separate (Supplementary Materials). The computed value, averaged over 5 different experiments, was $\theta_{eq} = (18 \pm 5)^o$. However, this should be adopted as an upper limit value, since local field corrections lead to an increase in the angle of the cone of attraction [31], and consequently to a decrease in the value of the computed equilibrium contact angle. The estimated value qualitatively matches with the small value reported by Dynal Biotech Technical Support, $\theta_{eq} = 10^o$ [32].

**Surface active additives.** As surface active agents we used anionic surfactant sodium dodecylsulphate (SDS, Fisher BioReagents, USA), with a water solubility of 130 g/L at 20°C, dissociation constant of 1.31 at 20°C, and critical micellar concentration (CMC) 8.2 mM, and the cationic surfactant Dodecyltrimethylammonium bromide (DTAB, Sigma Aldrich, USA), with a solubility in water of 954 g/L at 20°C (CMC = 13 mM). In all the experiments, the surfactants were used at a concentration well below their CMC.

3. Theory

**Field induced rotation of the superparamagnetic particles.** Magnetic particles upon the influence of a magnetic field $\boldsymbol{H}(\omega_f)$ rotating at an angular velocity $\omega_f$ are used as externally excitable probes in the study of the adsorption mechanism. The applied magnetic torque, given by $\tau_m = \mu_0 \langle m \times H(\omega_f) \rangle$, is determined both by the intensity and frequency of the field and by the size and spatial distribution of the magnetic grains. Here, $m$ is the particle's magnetic moment, $\mu_0 = 4\pi \cdot 10^{-7}$ NA$^{-2}$ is the magnetic permeability of the vacuum, and $\langle \cdots \rangle$ denotes a time average. When the rotating magnetic field is applied on "ideal" superparamagnetic particles, with very short Néel relaxation times as compared to the field frequency, the orientation of the induced magnetic dipole coincides almost instantaneously with that of the field, and the averaged applied magnetic torque is very small. However, when the micro-particles contain relatively large domains, with Néel times comparable to those of the inverse of the field frequency, then $\boldsymbol{m}$ is not instantaneously aligned with the rotating magnetic field, and the torque generated tends to align the particle's magnetic moment with the direction of the field [33].

At low Reynolds number, the magnetic torque is instantaneously balanced by the viscous torque $\tau_\eta(h) = 8\pi C_{\text{dragg}}(h)\eta a^3 \omega$, where $\omega$ is the angular velocity of the particle and $\eta$ the medium viscosity. The dimensionless parameter $C_{dragg}(h)$ is included to introduce the wall effect when the particle rotates in the vicinity of an interface. Here, $h$ is the distance between the center of the sphere and the confining plane (Figure 1). Extrapolating the data provided by Lee et al. [34] to the configuration where the sphere is just touching the interface, $C_{dragg}(a) = 1.15$.

At low field frequency, the particles rotate synchronously with the field, with a constant phase difference. As the frequency of the magnetic field increases, both the angular velocity of rotation of the particle and the phase increase. The field frequency for which the phase difference between the field and the magnetic moment is 90 degrees defines the critical frequency $f_c$. The value of $f_c$ is determined by the viscosity of the medium, the presence of neighbor boundaries, the field strength and the magnetization of the particles. For field frequencies $f > f_c$, the magnetic torque is not strong enough to compensate the viscous torque that the particle feels when rotating at the field frequency. In this regime, the particles asynchronously rotate in the same direction as the field, but at a lower frequency [33]. Within the bulk, the induced rotation is not converted into any translational motion, since the friction coefficient of the spherical particles is isotropic and there is no privileged direction along which the translation can occur [35].

**Roto-translational mechanism.** If the particle rotates with its axis of rotation parallel to a nearby flat boundary, the asymmetry introduced in the system favors their propulsion along the plane of the interface. This roto-translational coupling mechanism can be understood by considering that both the translational and rotational friction coefficients of the particle are functions of the distance to the interface [36]. In this process, the boundary conditions imposed by the interface are decisive in the transport mechanism. If the rotation of the particles is induced in the vicinity of a solid/liquid with stick boundary conditions, hydrodynamic interactions induce rolling motion. When the interface exhibits slipping boundary conditions, the particles are transported in the opposite direction, exhibiting an anti-rotational motion (sketch in Figure 1) [37]. The coupling between the angular and the linear velocity is given by $v = a\omega C'_{dragg}(h)$, where $C'_{dragg}(h)$ is a hydrodynamic parameter that increases with the proximity to the wall. The value of $C'_{dragg}(a) \approx 0.11$ was determined by measuring the ratio between the averaged maximum linear and angular velocities depicted by Janus magnetic particles -Dynabeads M270 semi-coated with a thin layer of gold- when propelled on the fluid interface through the field induced roto-translational mechanism. Since differences in the surface charge of the Janus spheres may have some effect on the roto-translational mechanism of the particle, experiments were also performed with permanently bonded dimers formed by the magnetic particles and the smaller non-magnetic silica spheres, 1 µm in diameter, reaching similar results (Figure S3 in Supplementary Materials).

**Interfacial torques.** Once adsorbed, the rotations of the particles around the axes tangent to the fluid interface are severely hindered (sketch in Figure 1). When the three-phase contact line is anchored to the surface of the trapped particle, the spinning of the particle around an axis parallel to the fluid interface causes deformation of the interface. The induced interfacial torque can be expressed, for small deformations, as [38]:

$$\tau_{int}(t) = \frac{96\gamma a^2}{\pi^3}\sin\phi(t)\sin\theta_{eq}\sin\frac{\Delta\theta_{eq}}{2} \qquad (3)$$

where $\Delta\theta_{eq} = \theta_A - \theta_R$ is the contact angle hysteresis, the difference between advancing $\theta_A$ and receding $\theta_R$ contact angles, and $\phi(t)$ the angle between the contact line and the vertical (sketch in Figure 1). The forced rotation of the adsorbed particle is also counteracted by the dissipative torque generated during the displacement of the three-phase contact line [19,23,26,29].

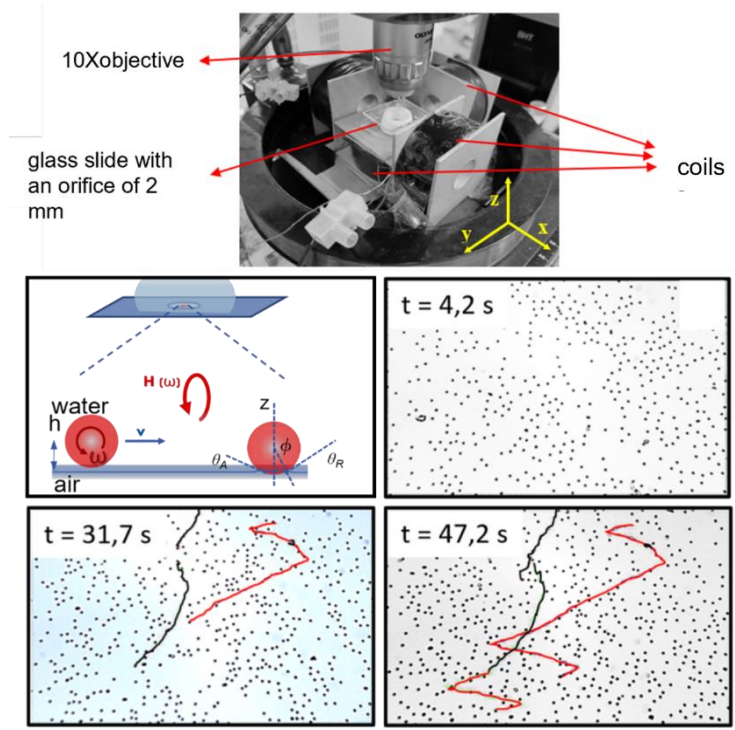

**Figure 1:** The picture shows the sample placed in the center of a set of coils made from insulated copper wire: a pair of coils oriented along the x-axis, facing each other at a distance equal to their diameter, and a third coil oriented along the z-axis, all placed in the proximities of the microscope stage. The diagram shows a drop of the aqueous dispersion held by capillary action on a glass slide with an orifice. After about 5 min, all the dispersed magnetic particles placed above the orifice are deposited by gravity on the slightly curved water-air interface, where they remain confined for the duration of the experiment. Particles are forced to rotate at an angular velocity $\omega$ and with their axis of rotation parallel to a nearby flat boundary. Non-adsorbed particles, floating a few nanometers above the fluid interface, exhibit coupled roto-translational motion, whereas in adsorbed particles the mechanism is strongly hindered by capillary torques. Here, $\theta_A$ and $\theta_R$ are the advancing and receding contact angles, respectively, and $\phi$ (t) the angle between the contact line and the vertical. The sequence of images shows the time behavior of the monolayer deposited on water/air interface when under the influence of an elliptically polarized rotating field, applied in the *xz* plane ($\beta$ = 0.87, $\mu_0H_0$ = 4.13 mT) (Movie 1). In these experiments, the rotational direction of the field is periodically changed to keep the particles in the field of view. For the sake of clarity, only three trajectories have been highlighted in the images (black lines: adsorbed particles show drift and Brownian motions drift motions, red lines: coupled roto-translational motion). One of the particles, initially adsorbed, desorbs and starts to show a field-induced motion after 32 s.

4. **Results** and discussions

**Degree of adsorption at the air-water interface and the colloidal stability of superparamagnetic particles in the presence of different additives.**

A series of experiments were carried out in which a drop of the corresponding solution (200 µL, 0,02 %w, in the presence of different additives at different concentrations) is deposited on the cell. After the 5 minutes necessary for all particles suspended in the aqueous phase to settle onto the water/air interface, where they float hundreds of nanometers above or are adsorbed (sketch in Figure 1), a rotating elliptically polarized field, of magnitude $\mu_0H_0$ = 4.13 mT, ellipticity $\beta$ = 0.87 and frequency $f$ = 20 Hz, is applied in the *xz* plane. Under these conditions, the low surface density of particles minimizes the role played by interparticle interactions in the adsorption process. In addition, the value of $\beta$ is chosen so that the interactions between adjacent particles, all them located in a plane parallel to the fluid interface, would be repulsive, thus minimizing the formation of dynamic colloidal assemblies, chains and/or disks, which "walk" and rotate along the fluid interface in a way that promotes the separation of their constituent particles from the boundary [39,40]. As the particles take a few minutes to settle to the bottom of the droplet, they arrive at a fluid interface loaded with the different additives. Therefore, in all these experiments the order of incorporation at the fluid interface is 1) additives 2) particles.

In these experiments, the field changes direction of rotation every 60 s, so that the particles trajectory can be followed for as long as possible, without particles leaving the field of view observed through the microscope (Movie 1, Figure 1). As already mentioned, the flow generated by the rotation of the particles floating on the fluid boundary induces an anti-roll motion, while the mechanism is gradually hindered by the interfacial torques when the particles enter the interface, where the adsorbed particles exhibit only drift motion. Hence, by monitoring the percentage of mobile particles it is possible to distinguish in real time the proportion of adsorbed particles.

In a parallel experiment, designed to characterize the colloidal stability in the presence of different additives at different concentrations, a constant magnetic field $\mu_0 H_{x0} = 3.6$ mT is applied parallel to the laden interface for ten minutes. After this time, the field is removed and the proportion of bonds that resist the absence of field is computed. In the absence of the applied field, the induced linear structures disintegrate due to 1) the superparamagnetic character of the particles, 2) the effect of the Brownian motion and 3) the electrostatic repulsion between the particles, unless particles form permanent bonds, induced by short-range van der Waals attractions [41,42]. The results obtained with these two techniques, together with those obtained with the mobility measurements, are presented below, when the particles adsorb in the presence of three different additives.

*Monovalent electrolyte: NaCl.* Figure 2.a,i shows the percentage of adsorbed particles measured 10 minutes after preparing the fluid interface, $\Gamma(10\text{min}) \times 100$, when the NaCl concentration changes from 0.1 to 100 mM. The percentage monotonically increases from $(43 \pm 17)$ %, when the experiments are performed in pure water, to $(100.0 \pm 0.5)$ %, when the salt concentration is greater than or equal to 1.0 mM, and presents a sharp transition over a range of salt concentration spanning hundreds of µM, between 600 and 1000 µM.

Under these diluted conditions, the adsorption percentage of 100% corresponds approximately to an adsorbed amount of 1 g/m$^2$.

Figure 2.a,ii shows that the degradation of colloidal stability is a gradual process, spanning more than two decades of electrolyte concentration. The onset of permanent bond formation and particle adsorption start at approximately the same range of electrolyte concentrations, between 600 and 1000 µM, which approximately coincides with the concentration in which the zeta potential $\zeta$ of the particles is below ± 10 mV (Figure 2.a,iii) –which is the threshold traditionally adopted to establish colloidal stability [43]. The electrolyte shields the negative surface charge of the particles and interface, lowering the potential barriers that the particle must overcome to form permanent interparticle bonds and to adsorb (Figure 3). In the absence of adsorption barrier, the interface behaves as a perfect sink [8].

*Cationic surfactant DTAB.* Figures 2.b,i, 2.b,ii and 2.b,iii show the results obtained by adding different concentrations of cationic (DTAB) and anionic (SDS) surfactants. According to Figure 2.b,i, the particle adsorption enhances with increasing DTAB concentration. The transition between the conditions where a relatively high fraction of particles is not adsorbed to those where all particles are adsorbed occurs in a narrower region, spanning tens of micromolar, at additive concentrations that are several orders of magnitude lower as compared to the one detected in the experiments performed with NaCl. The formation of permanent aggregates is strongly promoted when the DTAB concentration changes from 0.01 to 0.1 CMC (Figure 2.b,ii). Above 0.1 CMC, the mechanism of formation of permanent bonds, probably based on the emergence of capillary or hydrophobic attractions, which appear as a result of the minimization of free energy due to the unfavorable nature of the contact between the hydrophobic tails and the aqueous solvent [44], is completely efficient and large superparamagnetic pearl-chain-like structures are formed (Movie 2 and Figure 3). It is important to note that there exists a concentration range of DTAB, between 0.005 to 0.01 CMC, in which all particles adsorb at the interface and only a small proportion form permanent structures [45]. In this region, the surfactant decreases the adsorption barrier of particles at the air-water interface but without significantly affecting the wetting and electrostatic properties of particles. Hence, they remained stabilized mostly through electrostatic repulsions operating in water [45]. Above 0.05 CMC, the cationic surfactant adsorbs onto oppositely charged particles, and the surface charge of the particles is neutralized. The absolute value of the $\zeta$ potential reaches values below 10 mV (Figure 2.b,iii), and at the same time their hydrophobicity increases [46]. At higher concentrations, we do not observe any charge reversal phenomena via bilayer formation [21].

*Anionic surfactant, SDS.* In the presence of the anionic surfactant SDS, the trend is the opposite of the previous two cases. The percentage of adsorbed particles gradually varies from 40%, at a concentration of 0 µM, to 9.5%, at a concentration of 0.61 CMC (Figure 2.b,i). Besides, the presence of the anionic surfactant inhibits the formation of permanent bonds, which changes from 10 to roughly 0 % (Figure 2.b,ii). Since the increase in the SDS concentration does not appreciably enhance the particle mobility (Figure 2.b,iii), the improved colloidal stability is explained in terms of the competition for interfacial adsorption between the microparticles and the equally charged surfactants [47]. In the experimental setup used, it is not possible to measure at higher SDS concentrations, as the surface tension is so low that the pending droplets fall.

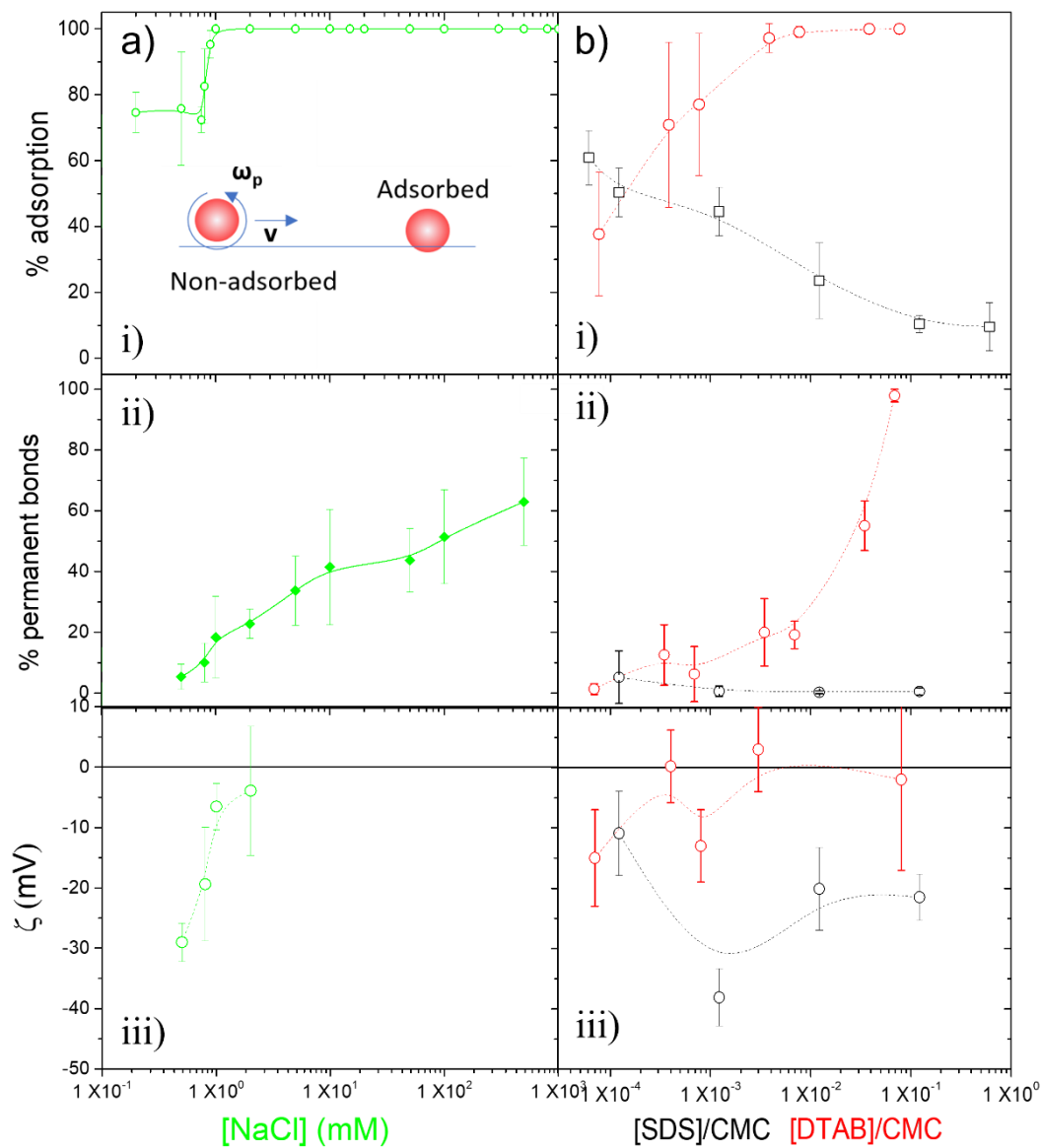

**Figure 2:** a, i) and b, i). Percentage of adsorbed particles as a function of NaCl, DTAB and SDS concentration. Data were assessed 10 minutes after depositing the droplet on the slide, once all the particles were in the proximity of the interface. a, ii) and b, ii). Percentage of permanent bonds formed between superparamagnetic particles, as a function of NaCl, DTAB and SDS concentration, after application of a constant field ($\mu_0 H_0$ = 3.6 mT), oriented parallel to the interface, for 10 min. a, iii) and b, iii). Variation of the ζ of as a function of the NaCl, DTAB and SDS concentrations. Lines are a guide for the eyes.

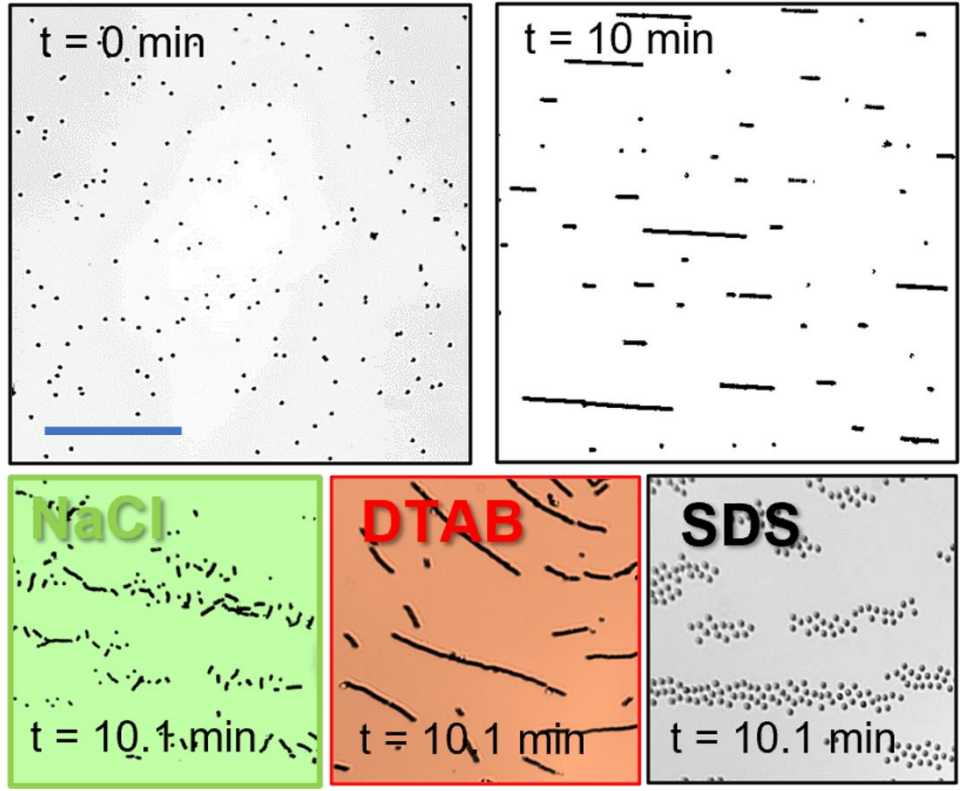

**Figure 3:** The images show the initial state of the particles at the interface, the field-induced chains formed after 10 minutes, and the partial or total disintegration of the linear chains after field removal, for concentration values [NaCl] = 500 mM, [DTAB] = 0.06 CMC and [SDS] = 0.012 CMC. In these experiments, the field $\mu_0 H_0$ = 3.6 mT was oriented parallel to the interface. Scale bar: 75 μm.

Figure 4 shows the velocity histograms constructed by tracking of the trajectories of a set of particles binned into 30 equally spaced bins. The analysis was carried out at different salt and surfactants concentrations, 10 minutes after the formation of the pendant drop, when adsorbed and non-adsorbed particles are placed close to the fluid interface. The histogram shows a bimodal distribution, indicating the presence of two well defined populations of particles. We adjusted the distribution to a sum of two Gaussian functions

$$y = y_0 + (A_1/(W_1\sqrt{(\pi/2)}))exp(-2((v-v_{c1})/W_1)^2) + (A_2/(W_2\sqrt{(\pi/2)}))exp(-2((v-v_{c2})/W_2)^2)$$

, which exhibits qualitative agreement with the experimental results. Here $W_1$ and $W_2$ are the respective variances and $A_1$ and $A_2$ two normalizer parameters. To disregard the collective convective motions, difficult to eradicate experimentally at fluid interfaces, the first peak of the fitted curves and experimental results were centered a posteriori at $v = 0$ µms$^{-1}$, after adding or subtracting tenths of µm·s$^{-1}$ to. After including the previous correction, the value of the velocity of the particles exhibiting roto-translational coupling is mostly ranged between 0.2 and 0.8 µm·s$^{-1}$ for all systems and concentrations. Figure 2 shows that the variation in the surfactant concentrations changes the proportion of adsorbed particles but apparently has no influence on the position of the peak of the Gaussians, nor in their variance. The addition of salt appears to decrease the averaged velocity of the rolling spheres and increase the variance of the Gaussians.

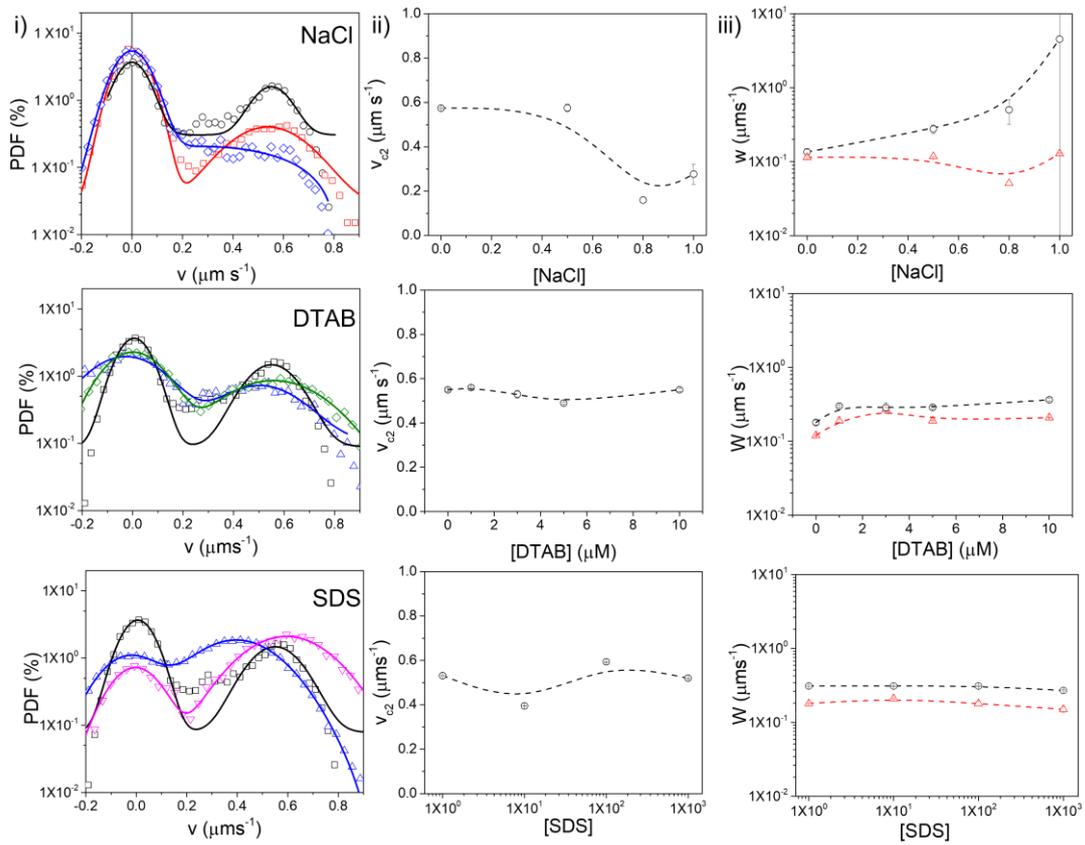

**Figure 4:** i) Normalized frequency distribution of the absolute value of the particle velocities, measured for different NaCl concentrations (black circles 0.0 mM, red squares 0.5 mM and blue diamonds 0.8 mM), different DTAB concentrations (black squares 0.0 µM, green diamonds 3.0 µM and blue upward triangles 10.0 µM) and different SDS concentrations (black squares 0.0 µM, blue upward triangles 10.0 µM and pink downward triangles 1000.0 µM). The continuous lines show the deconvolution of the data into two

Gaussians. ii) Position of the peak of the second Gaussian for different NaCl, DTAB and SDS concentrations. iii) Variances $W_1$ (red triangles) and $W_2$ (black circles) for different NaCl, DTAB and SDS concentrations. The dotted lines are guide to the eyes.

**Monitoring the particle adsorption/desorption processes.** In addition to determining the fraction of adsorbed particles, the proposed method allows to follow the dynamics of adsorption. Figure 5a) shows the time evolution of the velocity of 5 anti-rolling microspheres adsorbing at a bare water/air fluid interface (Movie 3).

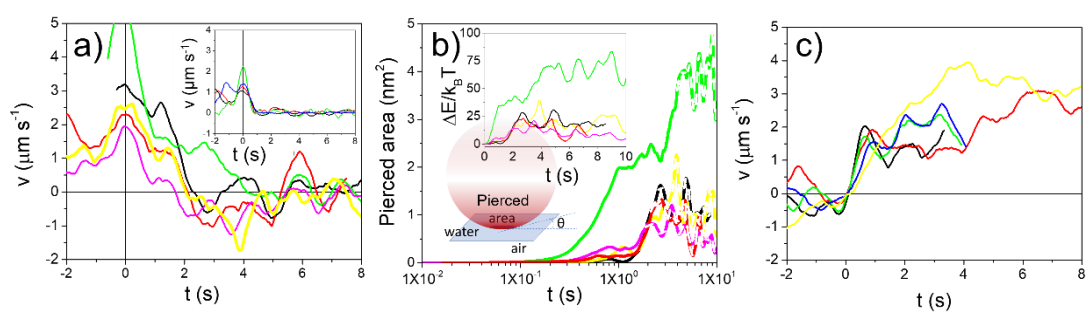

**Figure 5.** a) Slowing down of 5 non-adsorbed particles after adsorption at the air/water interface, at t = 0s. In the inset, the 5 different particles adsorb at a glass/water solid interface. b) Time evolution of the area pierced by the rotating spheres (continuous lines), which is accessible only before the particles are completely arrested by the action of the interfacial torque (dashed lines). The inset shows the corresponding attachment energies. All these curves were computed from the application of Equation 4 to data in a). c) Acceleration of 5 adsorbed particles after desorbing from the air/water interface, at t = 0s. In all the experiments, the area fraction of the particle monolayer was 1%.

Prior to adsorption, the particles are propelled on the fluid interface through the coupled roto-translational mechanism. During the first stage of the adsorption process, -1.0 s < t < 0.0 s in Figure 5a, particles rolling on the surface gradually increase their velocity over a time interval in which they approach the interface, so that coupled roto-translational mechanism progressively increases its efficiency. When the rotating sphere is just touching the interface, the applied magnetic torque is only balanced by the viscous torque, and $\tau_m = \tau_\eta(a) = \dfrac{C_{\text{dragg}}(a)}{C'_{\text{dragg}}(a)}(8\pi\eta a^2 v)$. According to the previous expression, the magnetic torque applied in the experiments ranges between $8\cdot 10^{-19}$ and $2.8\cdot 10^{-18}$ N·m$^{-1}$, close to the one reported in the literature [48]. The measured dispersion is mostly due to the heterogenous magnetic content of the particles [33]. Afterwards, the particles gradually reduce their velocity in the range between [-0.5, 0.5] µm·s$^{-1}$, reaching the adsorbed state. Similar adsorption processes have been detected in the case of particles floating on a glass surface (inset in Figure 5a). However, the velocity of the particles adsorbing at the fluid interface shows higher fluctuations, which can be attributed to a more significant effect of convection. We can quantify the dynamics of adsorption using the time it takes for the particles to stop when they reach the interface. With this measure, the dynamics of adsorption at the fluid interface is significantly slower (adsorption lasts several seconds) than at the solid interface (adsorption occurs in less than one second). At fluid interfaces, the applied magnetic torque is gradually compensated by the interfacial torque during these time intervals:

$$\tau_m - \tau_I = \tau_\eta(a) \qquad (4).$$

During adsorption, the magnetic torque remains constant, the viscous torque decreases with decreasing particle velocity, while the interfacial torque, described by Equation 3 and obtained from the difference between the magnetic and the viscous torques, increases with the degree of particle penetration. Hence, the tracking of particles' deceleration allows to follow the time evolution of $\phi$ during the very earliest stage of the adsorbing process (continuous lines in Figure 5b), before the magnetic torque is completely balanced by the capillary torque (dashed lines in Figure 5b). Figure 5b shows that the acceleration and deceleration processes differ between different particles, as well as the velocity changes occur when the area of the perforated interface is less than 5 nm$^2$. The uncertainty in the pierced area, mainly determined by the uncertainty in the radii and velocity of the particles, is about 1 nm$^2$. As shown in the inset in Figure 5b, the pierced area corresponds to an attachment energy smaller than 75 k$_B$T and to changes in the vertical position of less than one nm, not detectable by means of holographic techniques [17]. It should be noted that the contribution of the interfacial dissipative torque, generated by the contact line friction, is neglected in these calculations, so that the values of both the interfacial torque and the pierced area are systematically overestimated. The dissipative torque may be significant during this initial period, when particles are far from the equilibrium state, and both the angular velocity of the particles and the adsorption rate are higher [48].

During adsorption, at least some particles exhibit slow upward and downward motions, moving in and out of the gas phase, in a process governed by capillary forces and the hopping over surface defects. The upward movement against the capillary force is possible because the equilibrium contact angle of these particles is small, so before arriving to the equilibrium position the particles pass through configurations in which the binding energy is close to the thermal energy. In fact, a small proportion of the particles spontaneously desorb, contrary what is observed in most of the experimental systems, where the equilibrium contact angle is typically higher [17,27]. Figure 5c shows the time evolution of the velocity of 5 different particles, when they pass from the adsorbed to the non-adsorbed anti-rolling state. Most of the velocity changes are detected during the first second after desorption, so the comparison between the time evolutions depicted in Figures 5a and 5c shows that the later stages of the desorption process occur faster than the earlier adsorption stages.

**Adsorption dynamics.** The time evolution of the fraction of adsorbed particles $\Gamma(t)$ was measured, for two hours, as a function of the salt concentration (Figure 6). Monitoring the percentage of adsorbed particles began seconds after the creation of the fluid interface, even when not all particles had settled.

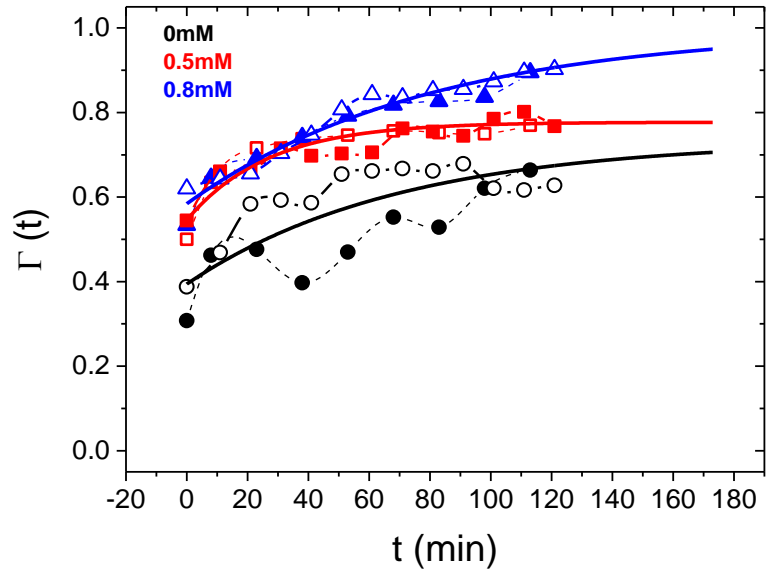

**Figure 6:** Variation of the fraction of adsorbed particles $\Gamma$(t), at different NaCl concentrations (black circles 0.0 mM, red squares 0.5 mM, and blue triangles 0.8 mM), as a function of time. The data represented with open symbols were measured under the uninterrupted application of the rotating field, while for the solid points the field was imposed for 30 seconds every 15 minutes. The continuous lines are the fits of Equation 5 to the experimental data. Measurement errors have not been included so that the results can be interpreted more clearly. Dashed lines are a guide for the eye.

For the all the imposed NaCl concentrations, the measured percentage of adsorbed particles presents a relatively high initial value and a subsequent gradual growth with time. At surfactant concentrations higher than 2 mM, results not shown, the particles adsorbed almost immediately and irreversibly. In dilution conditions, all settled particles float within hundreds of nanometers of the fluid interface and particles do not compete for adsorption. In these conditions the change in adsorption percentage is proportional to the proportion of non-adsorbed particles, 1-$\Gamma$(t), and the adsorption constant $k_{ads}$. Similarly, the kinetics of the desorption process is proportional to $\Gamma$(t) and the desorption constant $k_{des}$, so $\frac{d\Gamma}{dt} = k_{ads}(1-\Gamma(t)) - k_{des}\Gamma(t)$. The above equation leads to an exponential relaxation, on the time scale $\tau^{-1} = k_{ads} + k_{des}$, given by:

$$\Gamma(t) = \left( \frac{k_{ads}}{k_{ads}+k_{des}}(1-e^{-(k_{ads}+k_{des})t}) + \Gamma(0)e^{-(k_{ads}+k_{des})t} \right) \quad (5)$$

where $\Gamma(0)$ is the initial fraction of adsorbed particles. From the fit of Equation 5 to the experimental values, we find the parameter values shown in Table 1. Here, $\Lambda$ is the ratio between the characteristic settling and adsorption times, and $L$ is the typical height of the pending droplet (approx. 3 mm).

| [NaCl] (mM) | $k_{ads}^{-1}$ (min) | $k_{des}^{-1}$ (min) | $k_{des}/k_{ads}$ | $\tau$ (min) | $\Lambda = \frac{6\eta L(k_{ads}+k_{des})}{4/3\, a^2 g \Delta\rho}$ | $\Gamma(0)$ |
|---|---|---|---|---|---|---|
| 0 | 95 | 262 | 0.36 | 69.7 | 0.16 | 0.40 |
| 0.5 | 36 | 111 | 0.32 | 27.0 | 0.43 | 0.53 |
| 0.8 | 80 | 25000 | 0.0032 | 76.7 | 0.15 | 0.58 |

**Table 1:** Different adsorption parameters, defined throughout the text, as a function of the NaCl concentration.

As shown in Table 1, the value of the desorption constants increases with increasing salt concentration, while the ratio of the $k_{des}/k_{ads}$ constant values decreases. As the salt concentration increases the desorption process is less favorable and the particles adsorb irreversibly. The differences observed between the relatively high initial fraction of adsorbed particles $\Gamma(0)$ and the subsequent long adsorption time scales could be explained in terms of heterogeneity of wetting properties, probably derived from the distribution of particle surface properties [49].

These results were reproduced by means of two different types of experiments, designed to test whether the rotation of the magnetic particles influences the adsorption kinetics. The first experiment consists of the uninterrupted application of the rotating field (open symbols in Figure 6), while in the second experiment the same field is applied in 30-second pulses at 15-minute intervals (solid symbols in Figure 6). The previous data show that the trends collected in both types of experiments are comparable, which seems to indicate that the induced rotation of the particles, with an angular velocity of the order of about 30 rad/s, does not influence the adsorption dynamics of the particles.

**Conclusions**

In this work, we study the effect of salts and surfactants on the adsorption of micrometric Dynabeads M-270 superparamagnetic particles at the air-water interface. Our results show that the addition of a monovalent electrolyte promotes both the adsorption and the formation of permanent interparticle bonds. Alternatively, the cationic surfactant presents a narrower transition between the concentrations where a relatively high fraction of the particles is not adsorbed to those where all particles are adsorbed. In this case, the formation of permanent aggregates of particles is strongly promoted for cationic surfactant concentration larger than 0.1 CMC. On the other hand, the presence of an anionic surfactant inhibits both the adsorption of the particles and the formation of permanent bonds. We also characterize both the adsorption dynamics of individual particles in the early stages of the process, when the protruded interface area is less than 5 nm2 and the adsorption energy less than 75 kBT, and the spontaneous desorption of a small proportion of the particles, possible since the value of the equilibrium contact angle is small and the relaxation of the particle position towards equilibrium is slow [26].

The results were obtained using a new and simple strategy based on tracking the roto-translational motion of a set of monomers resting on a nearby boundary in the presence of different additives. In the experiments, superparamagnetic spherical colloids are forced to rotate under a rotating magnetic field in the vicinity of an air/water interface. Since the piercing of the interface by a particle favors the suppression of the roto-translational hydrodynamic mechanism, particle tracking allows a statistical treatment of the adsorption process, together with the monitoring of the dynamics of the adsorption/desorption processes.

Unlike classical tensiometry techniques, which require relating the measured evolution of surface pressure to surface coverage through an appropriate model [2,6,8,9], and which are strongly determined by the effect that bulk transport, diffusion or concentration may have on the adsorption mechanism [7,50], the proposed strategy allows direct and simple measurement of adsorption and desorption kinetics. Under conditions where the particles have to overcome the electrostatic energy barrier to adsorb, both the adsorption and desorption rate constants are in the order of $10^{-4}$ $s^{-1}$, larger to those estimated in the competitive adsorption of surfactants and of gold nanoparticles, at relatively high particle concentrations [50], similar to those assessed, using a generalized Frumkin model, in the adsorption of nonionic surfactants onto a clean air-water interface [51,52], or to those reported in the desorption proceeding under barrier control of decanol and dodecanol monolayers spread in pure water, at several surface pressures and temperatures [53].

On the contrary, the strategy proposed for the study of adsorption dynamics is indirect, based on the evolution of the particle velocity, approximate, since it neglects the torque due to the fluctuation of the contact line, and limited to the first instants of the adsorption process, when the applied torque is greater than the interfacial torque. In its favor, however, we must emphasize that this methodology does not require the complex optical configurations and analysis tools used by interferometric techniques [26-29]. Furthermore, since the applied magnetic torque is only able to overcome the interfacial torque caused by interface deformation during the initial phase of the adsorption process, this approach opens the door to assess the time evolution of the vertical position of the protruding particles with resolutions below the nanometer range, below what is achieved in the aforementioned techniques.

The described method can be easily generalized to other rotating probes, including particles subjected to rotating electric [54], acoustic [55] or optical fields [56], made up of different materials, coated with different molecules, polydisperse and adsorbing on other solid or fluid interfaces, loaded with different surfactants (nonionic surfactants, proteins…). In particular, the use of particles with higher magnetic moments (larger particles, ferromagnetic particles or paramagnetic particles with higher susceptibilities), or the application of stronger rotating fields, could improve the resolution in the monitoring of the particle protrusion during the first stages of the adsorption mechanism.

**Acknowledgments**

We would like to Eduardo Guzmán for fruitful discussions. This work was partially funded by Horizon 2020 program through 766972-FET-OPEN21 NANOPHLOW and Ministerio de Ciencia e Innovacion (Grants No. PID2019-105343GB-I00 and PID2019-106557GB-C21). F.M.-P. acknowledges support from MINECO (Grant No. RYC-2015-18495).